\documentstyle[12pt]{article}
\def\CH{{\cal H}}
\def\CBPi{\bar{\cal P}{}^\prime}
\def\CBPii{\bar{\cal P}{}^{\prime\prime}}
\def\CBP{\bar{\cal P}}
\def\CP{{\cal P}}
\def\CO{{\cal O}}

\begin{document}

\thispagestyle{empty}

\baselineskip=0.6cm

\noindent P.~N.~Lebedev Physical Institute              \hfill
Preprint FIAN/TD/8--94\\ I.~E.~Tamm Theory Department       \hfill
\begin{flushright}{August 1994}\end{flushright}

\begin{center}

\vspace{0.5in}

{\Large\bf SPLIT INVOLUTION COUPLED TO }

\bigskip

\vspace{0.41cm}

{\Large\bf ACTUAL GAUGE SYMMETRY}

\bigskip

\vspace{0.3in}

{\large  I. A. Batalin$^1$, S.L.Lyakhovich$^2$ and I. V. Tyutin$^1$}\\
\medskip

$^{(1)}${\it P.N.Lebedev Physical Institute, 117924 Moscow, Russia}$^{\dagger}$

$^{(2)}${\it Tomsk State University, 634050 Tomsk, Russia}

\end{center}

\vspace{1.5cm}

\centerline{\bf ABSTRACT}

\begin{quotation}

The split involution quantization scheme, proposed previously for
pure second--class constraints only, is extended to cover the case
of the presence of irreducible first--class constraints. The explicit
Sp(2)--symmetry property of the formalism is retained to hold. The
constraint algebra generating equations are formulated and the Unitarizing
Hamiltonian is constructed. Physical operators and states are defined in the
sense of the new equivalence criterion that is a natural counterpart to the
Dirac's weak equality concept as applied to the first--class quantities.

\end{quotation}

\vfill

\noindent

$^{\dagger}$ E-mail address: batalin@fian.free.net,tyutin@fian.free.net

\newpage

\setcounter{page}{2}

\section{Introduction}

In previous paper [1] of the present authors the split involution formalism
has been proposed for canonical quantization of dynamical systems with pure
second--class constraints.

The formalism implies no extra variables to be introduced with the purpose
of converting original constraints into effective ones of the first-class.

On the other hand, the total set of original second--class constraints is
supposed to be polarized by splitting into two interchangeable subsets,
$T^a_\mu$, $a=1,2$, to satisfy the so-called "split involution" relations

$$
(\imath\hbar)^{-1}[T^{\{a}_\mu,T^{b\}}_\nu]=
U^{\{a\rho}_{\mu\nu}T^{b\}}_\rho,
\eqno{(1.1)}$$
symmetrized in their superscripts $a$, $b$.

Besides, the Hamiltonian $H$ is supposed to satisfy the relations
\footnote{It goes without saying that arbitrary second--class constraints
(whose Fermionic component number is divisible by 4) and Hamiltonian
can be transformed locally to the polarized basis subjected to eqs. (1.1),
(1.2). What is not so evident that there exists a valuable set of relativistic
dynamical systems such that the Dirac's hamiltonianization procedure, being
applied directly to the original relativistic Lagrangian, just produces the
polarized constraint basis.}

$$
(\imath\hbar)^{-1}[H,T^a_\mu]=V^\nu_\mu T^a_\nu.
\eqno{(1.2)}$$

One generates the "gauge" algebra, initiated by the relations (1.1), (1.2)
by solving the equations

$$
[Q^a,Q^b]=0,\quad[Q^a,\CH]=0,
\eqno{(1.3)}$$
for the Fermions $Q^a$ and Boson $\CH$ in the form of a series expansion
in ghost powers

$$
Q^a=C^\mu T^a_\mu+\ldots,\quad\CH=H+\ldots.
\eqno{(1.4)}$$
Then one constructs the complete Unitarizing Hamiltonian of the theory in the
following Sp(2)--symmetric form

$$
H_{complete}=\CH+\varepsilon_{ab}(\imath\hbar)^{-2}[Q^b,[Q^a,B]]
\eqno{(1.5)}$$
where $B$ is a "gauge--fixing" Bosonic operator. Being the physical quantities
defined in an appropriate way, they do not depend on a particular choice of
a ``gauge'' operator $B$. This independence is quite a nontrivial feature of
the split involution scheme, because pure second--class constraints do not
generate an actual gauge symmetry.

The algebra generating equations (1.3) as well as the Hamiltonian (1.5) possess
the $Sp(2)$--covariant form which is characteristic to the formalism developed
in Refs. [2, 3] to quantize gauge--invariant theories in a ghost--antighost
symmetric fashion. However, the number of ghosts (and antighosts) introduced
in the formalism [2, 3] is twice as compared with the corresponding number in
the split involution theory.  Moreover, the ghost numbers of the generating
operators ($Q^1$, $Q^2$) are ($+1$, $+1$) in the split involution scheme,
while in the ghost--antighost symmetric theory these numbers are ($+1$, $-1$).

In the present paper we generalize the split involution formalism by including
original first--class constraints into it. When doing this we retain the
explicit $Sp(2)$--symmetry property of the method to hold.

We assign ghost canonical pairs to constraints of both the classes and require
the ghost number operators $G^\prime$ and $G^{\prime\prime}$ of the
first and second class, respectively, to be conserved separately. In accordance
with this requirement, a pair of the ghost number values, denoted by
$\hbox{gh}^\prime$ and $\hbox{gh}^{\prime\prime}$, is assigned to each
admitted operator of the theory.

Then we formulate the extended version of the gauge algebra
generating equations. We require the generating operator of the first--class
constraint algebra to be nilpotent modulo contributions similar to the
gauge--fixing term in r.h.s. of (1.5). Thereby we define the equivalence
criterion that is a natural counterpart to the Dirac's weak equality concept as
applied to the first--class quantities. The conservation property of the
first--class generating operator is also formulated in the sense of the new
equivalence criterion proposed.

The constraint algebra generating equations are shown to possess the group of
automorphisms that enables one to make the first (resp. second)--class
constraints be a set of momenta (resp. a set of canonical pairs). The maximal
group of automorphisms is given by semidirect product of three groups that are:
ghost--dependent canonical transformations,$c$--numerical symplectomorphisms,
and exact shifts initiated by the new equivalence criterion.

In terms of the constraint algebra generating operators we construct the
complete Unitarizing Hamiltonian of the theory. We modify the definition (1.5)
by adding the genuine gauge-fixing term required by the presence of original
first--class constraints.

Finally, we formulate the definitions of physical operators and physical
states in the sense of the new equivalence criterion.

{\bf Notations and Conventions.} As usual, $\varepsilon(A)$ represents the
Grassmann parity of the quantity $A$.

If $n=n_++n_-$ is the total number of some superobjects, then $n_+(n_-)$
indicates the number of Bosons (Fermions) among them.

The standard supercommutator of the operators $A$, $B$ is defined by the
formula
$$
[A,B]\equiv AB-BA(-1)^{\varepsilon(A)\varepsilon(B)}.
\eqno{(1.6)}$$

By $\varepsilon^{ab}$ we denote the constant $Sp(2)$--invariant tensor

$$
\varepsilon^{ab}=\left(\begin{array}{rc}0&1\\-1&0\end{array}\right),
\eqno{(1.7)}$$
while its inverse is denoted as $\varepsilon_{ab}$:

$$
\varepsilon^{ab}\varepsilon_{bc}=\delta^a_c.
\eqno{(1.8)}$$

We also use the standard notations for symmetrization

$$
A^{\{ab\}}\equiv A^{ab}+A^{ba},
\eqno{(1.9)}$$
and antisymmetrization

$$
A^{[ab]}\equiv A^{ab}-A^{ba}.
\eqno{(1.10)}$$

Greek indices of first(second)--class constraints are taken from the
first(second),half of the Greek alphabet,
$\alpha,\ldots,\lambda$($\mu,\ldots,\omega$). The same convention holds for
related quantities.

By $\hbox{gh}^\prime(A)$ ($\hbox{gh}^{\prime\prime}(A)$) we denote the
first(second)--class ghost number of quantity $A$.

The other notation is clear from the context.

\section{Constraint Algebra}

Let

$$
(q^i, p_i), \quad i=1,\ldots, n=n_++n_- \quad, \eqno{(2.1)}$$

$$
\varepsilon(q^i)=\varepsilon(p_i)\equiv\varepsilon_i,\quad
\hbox{gh}^\prime(q^i)=-\hbox{gh}^\prime(p_i)=0,\quad
\hbox{gh}^{\prime\prime}(q^i)=-\hbox{gh}^{\prime\prime}(p_i)=0,
\eqno{(2.2)}$$

$$
(q^i)^\dagger=q^i,\quad(p_i)^\dagger=p_i(-1)^{\varepsilon_i},
\eqno{(2.3)}$$
be a set of the original phase variable operators whose equal--time nonzero
super--commuta\-tion relations are

$$
(\imath\hbar)^{-1}[q^i.p_j]=\delta^i_j.
\eqno{(2.4)}$$
Further let us suppose the Hamiltonian,

$$
H=H(p,q), \quad\varepsilon(H)=0,
\eqno{(2.5)}$$
and the constraint operators,

$$
T_\alpha=T_\alpha(p,q),\quad\varepsilon(T_\alpha)\equiv
\tilde{\varepsilon}_\alpha,
\eqno{(2.6)}$$

$$
\alpha=1,\ldots,m^\prime=m^\prime_++m^\prime_-,
\eqno{(2.7)}$$

$$
T^a_\mu=T^a_\mu(p,q),\quad\varepsilon(T^a_\mu)\equiv\varepsilon_\mu,
\eqno{(2.8)}$$

$$
a=1,2;\quad\mu=1,\ldots,m^{\prime\prime}=m^{\prime\prime}_++m^{\prime\prime}_-,
\eqno{(2.9)}$$

$$
m^{\prime\prime}_-=2k,\quad m_\pm\equiv
m^\prime_\pm+m^{\prime\prime}_\pm<n_\pm,
\eqno{(2.10)}$$
to satisfy the following involution relations

$$
(\imath\hbar)^{-1}[T^{\{a}_\mu,T^{b\}}_\nu]=U^{\{a\rho}_{\mu\nu}T^{b\}}_\rho,
\eqno{(2.11)}$$

$$
(\imath\hbar)^{-1}[T^a_\mu,T_\alpha]=\tilde{U}^{a\beta}_{\mu\alpha}T_\beta+
U^\nu_{\mu\alpha}T^a_\nu,
\eqno{(2.12)}$$

$$
(\imath\hbar)^{-1}[T_\alpha,T_\beta]=\tilde{U}^\gamma_{\alpha\beta}T_\gamma+
{1\over2}\varepsilon_{ab}W^{\mu\nu}_{\alpha\beta}(T^b_\nu\delta^\rho_\mu-
T^b_\mu\delta^\rho_\nu(-1)^{\varepsilon_\mu\varepsilon_\nu}-
\imath\hbar U^{b\rho}_{\nu\mu})T^a_\rho,
\eqno{(2.13)}$$

$$
(\imath\hbar)^{-1}[H,T^a_\mu]=V^\nu_\mu T^a_\nu,
\eqno{(2.14)}$$

$$
(\imath\hbar)^{-1}[H,T_\alpha]=\tilde{V}^\beta_\alpha T_\beta+
{1\over2}\varepsilon_{ab}W^{\mu\nu}_\alpha(T^b_\nu\delta^\rho_\mu-
T^b_\mu\delta^\rho_\nu(-1)^{\varepsilon_\mu\varepsilon_\nu}-
\imath\hbar U^{b\rho}_{\nu\mu})T^a_\rho.
\eqno{(2.15)}$$
where the structure coefficient operators are some functions of the
original phase variables (2.1), and the following antisymmetry properties
are supposed to hold:

$$
U^{a\rho}_{\mu\nu}=-U^{a\rho}_{\nu\mu}(-1)^{\varepsilon_\mu\varepsilon_\nu},
\quad\tilde{U}^\gamma_{\alpha\beta}=-\tilde{U}^\gamma_{\beta\alpha}
(-1)^{\tilde{\varepsilon}_\alpha\tilde{\varepsilon}_\beta},
\eqno{(2.16)}$$

$$
W^{\mu\nu}_{\alpha\beta}=-W^{\nu\mu}_{\alpha\beta}
(-1)^{\varepsilon_\mu\varepsilon_\nu}=-W^{\mu\nu}_{\beta\alpha}
(-1)^{\tilde{\varepsilon}_\alpha\tilde{\varepsilon}_\beta},
\eqno{(2.17)}$$

$$
W^{\mu\nu}_\alpha=-W^{\nu\mu}_\alpha(-1)^{\varepsilon_\mu\varepsilon_\nu}.
\eqno{(2.18)}$$

Let us also require the supercommutators

$$
\Delta^{ab}_{\mu\nu}\equiv(\imath\hbar)^{-1}[T^{[a}_\mu,T^{b]}_\nu],
\eqno{(2.19)}$$
enumerated by collective indices $(a,\mu)$, $(b,\nu)$, to form an
invertible operator--valued matrix:

$$
\Delta\quad\Rightarrow\quad\exists\quad\Delta^{-1}
\eqno{(2.20)}$$
This condition implies the constraints (2.8) to be of the second--class.

In their own turn the involution relations (2.12),(2.13) imply the constraints
(2.6) to be of the first class. Let us require for these constraints
to commute with the operators (2.1) to give an operator--valued
supermatrix whose invertible Bose--Bose and Fermi--Fermi blocks are of the
maximal possible sizes $m^\prime_+\times m^\prime_+$ and $m^\prime_-\times
m^\prime_-$, respectively, which requirement is an operator version to the
irreducibility condition.

As for the second-class constraints, they are irreducible due to the
condition (2.20).

The irreducibility property determines the quantum rules of "dividing by
constraints", i.e. characteristic form of the most general operator solution
to the basic set of homogeneous linear equations

$$
Z^\mu T^a_\mu+\tilde{Z}^{a\alpha}T_\alpha=0,
\eqno{(2.21)}$$

$$
Z^{\{a\mu}T^{b\}}_\mu+\tilde{Z}^{ab\alpha}T_\alpha=0,
\quad\tilde{Z}^{[ab]\alpha}=0,
\eqno{(2.22)}$$

$$
Z^{ab\mu}T^c_\mu+\hbox{cycle}(a,b,c)=0,\quad Z^{[ab]\mu}=0,
\eqno{(2.23)}$$

$$
Z^{\mu\nu}{1\over2}\varepsilon_{ab}(T^b_\nu\delta^\rho_\mu-
T^b_\mu\delta^\rho_\nu(-1)^{\varepsilon_\mu\varepsilon_\nu}-
\imath\hbar U^{b\rho}_{\nu\mu})T^a_\rho+\tilde{Z}^\alpha T_\alpha=0,\quad
Z^{\mu\nu}=-Z^{\nu\mu}(-1)^{\varepsilon_\mu\varepsilon_\nu},
\eqno{(2.24)}$$
which are obtained by applying the Jacobi identity to all the involution
relations (2.11) -- (2.15). In the Appendix these equations will be  considered
in more details.

It would be just desirable to avoid imposing further restrictions on
the constraint algebra (2.11) -- (2.15). Unfortunately, we are unable to
prevent such restrictions for the present. Therefore we have to impose the
following extra condition on the structure coefficients
$\tilde{U}^{a\beta}_{\mu\alpha}$ entering the cross--sector relation (2.12)
that involves constraints of the both classes:

$$\begin{array}{c}
(\imath\hbar)^{-1}[T^{\{a}_\mu,\tilde{U}^{b\}\beta}_{\nu\alpha}]-
(\imath\hbar)^{-1}[T^{\{a}_\nu,\tilde{U}^{b\}\beta}_{\mu\alpha}]
(-1)^{\varepsilon_\mu\varepsilon_\nu}-\tilde{U}^{\{a\gamma}_{\mu\alpha}
\tilde{U}^{b\}\beta}_{\nu\gamma}(-1)^{\varepsilon_\nu
(\tilde{\varepsilon}_\alpha+ \tilde{\varepsilon}_\gamma)}+\\[9pt]
+\tilde{U}^{\{a\gamma}_{\nu\alpha}\tilde{U}^{b\}\beta}_{\mu\gamma}
(-1)^{\varepsilon_\mu(\tilde{\varepsilon}_\alpha+\tilde{\varepsilon}_\gamma+
\varepsilon_\nu)}-U^{\{a\rho}_{\mu\nu}\tilde{U}^{b\}\beta}_{\rho\alpha}=
\tilde{U}^{\{a\rho\gamma}_{\mu\nu\alpha}(T^{b\}}_\rho\delta^\beta_\gamma-
\imath\hbar\tilde{U}^{b\}\beta}_{\rho\gamma})
(-1)^{\varepsilon_\mu\tilde{\varepsilon}_\alpha},
\end{array} \eqno{(2.25)}$$
where the new structure coefficient operators
$\tilde{U}^{a\gamma\rho}_{\mu\nu\alpha}$ are supposed to possess the
antisymmetry property

$$
\tilde{U}^{a\gamma\rho}_{\mu\nu\alpha}=-\tilde{U}^{a\gamma\rho}_{\nu\mu\alpha}
(-1)^{\varepsilon_\mu\varepsilon_\nu+\varepsilon_\nu
\tilde{\varepsilon}_\alpha+\tilde{\varepsilon}_\alpha\varepsilon_\mu}.
\eqno{(2.26)}$$

Let us consider the status of the restriction (2.25). By applying the Jacoby
identity to the constraint algebra (2.11) -- (2.15) and then making use of the
above mentioned quantum "rules of dividing by constraints", one can show the
operators $\tilde{U}^{a\beta}_{\mu\alpha}$ to satisfy the relation that
differs from the one (2.25) by the extra contribution

$$
\tilde{\tilde{U}}{}^{ab\gamma\lambda}_{\mu\alpha}(T_\lambda\delta^\beta_\gamma-
T_\gamma\delta^\beta_\lambda
(-1)^{\tilde{\varepsilon}_\gamma\tilde{\varepsilon}_\lambda}-
\imath\hbar\tilde{U}^\beta_{\gamma\lambda})
\eqno{(2.27)}$$
to r.h.s. Thus, in fact, the condition (2.25) is equivalent to
the requirement for the contribution (2.27) to vanish.

On the other hand, one can consider the cross--sector relation (2.12) to be
the covariant constancy property of the constraints, being the structure
coefficients $\tilde{U}^{a\beta}_{\mu\alpha}$, $U^{a\rho}_{\mu\nu}$ treated to
serve as the connection components. From this viewpoint, l.h.s. of (2.25) is
nothing else but the corresponding curvature components. The condition
(2.25), being treated classically, requires for the curvature to vanish on
the second--class constraint surface, while the algebra (2.11) -- (2.13)
itself implies a weaker condition to be satisfied that the curvature
components should vanish on the surface of all the constraints.

Now let us comment in brief the most characteristic features of the involution
relations (2.11) -- (2.15).

First of all we observe that the split
involution relations (2.11), (2.14) retain their original form [1] specific to
the pure second--class constraint case. Further, the cross--sector constraint
supercommutators are actually restricted in two respects: the operators
$\tilde{U}^{a\beta}_{\mu\nu}$ are subordinated to the relations (2.25), and the
operators $U^\nu_{\mu\alpha}$ do not possess their own $Sp(2)$--indices.

Finally, let us turn to the first--class constraint involution relations
(2.13), (2.15). Being these relations treated classically, second--class
constraints are allowed to contribute only quadratically, which assertion is
a consequence of the Jacoby identity. Such quadratic contributions are just
represented by the second and third terms in r.h.s. of (2.13), (2.15), and
these terms possess the specific structure characterized by the antisymmetry
property of the coefficients $\varepsilon_{ab}W^{\nu\mu}$ in their indices
$a$, $b$ and $\mu$, $\nu$. However, at $\hbar\neq0$ second--class
constraints appear to be allowed quantum--mechanically to contribute to
(2.13), (2.15) linearly with the effective coefficients
$-{1\over2}\imath\hbar\varepsilon_{ab}W^{\nu\mu}U^{b\rho}_{\mu\nu}$. These
linear quantum contributions, represented by the fourth terms in r.h.s. of
(2.13), (2.15), are necessary in order to provide the operator compatibility
of the formal constraint algebra.

Given the initial operators (2.5), (2.6), (2.8), the involution relations
(2.11) -- (2.15) serve to determine the lowest structure coefficient operators

$$
U^{a\rho}_{\mu\nu},\quad\tilde{U}^{a\beta}_{\mu\alpha},\quad
U^\nu_{\mu\alpha},\quad \tilde{U}^\gamma_{\alpha\beta},\quad
W^{\mu\nu}_{\alpha\beta},\quad V^\nu_\mu,\quad \tilde{V}^\beta_\alpha,\quad
W^{\mu\nu}_\alpha
\eqno{(2.28)}$$
up to a natural arbitrariness.

By making use of the Jacoby identity together with the irreducibility property
of the constraints, one can derive the necessary compatibility conditions to
the involution relations (2.11) -- (2.15). These new conditions, including the
one (2.25), contain new structure coefficient operators to be determined at
this level. On the other hand, these relations reduce to an admissible extent
the arbitrariness in the preceding--level structure coefficient operators.
Continuing this procedure, one generates, step by step, an infinite gauge
algebra initiated by the operators (2.5), (2.6), (2.8).

In the next Section we formulate the generating equations that give
automatically an infinite set of structure relations of the constraint
gauge algebra.

\section{Constraint algebra generating equations}

As a next step let us introduce the ghost phase variable operators.
We assign a ghost canonical pair to each first--class constraint
operator:

$$
T_\alpha\quad\rightarrow\quad(C^{\prime\alpha},\CBPi_\alpha),\quad
\alpha=1,\ldots,m^\prime,
\eqno(3.1)$$

$$
\varepsilon(C^{\prime\alpha})=\varepsilon(\CBPi_\alpha)=
\tilde{\varepsilon}_\alpha+1,
\eqno{(3.2)}$$

$$
\hbox{gh}^\prime(C^{\prime\alpha})=-\hbox{gh}^\prime(\CBPi_\alpha)=1,\quad
\hbox{gh}^{\prime\prime}(C^{\prime\alpha})=-\hbox{gh}^{\prime\prime}
(\CBPi_\alpha)=0.
\eqno{(3.3)}$$

$$
(C^{\prime\alpha})^\dagger=C^{\prime\alpha},\quad(\CBPi_\alpha)^\dagger=
-\CBPi_\alpha(-1)^{\tilde{\varepsilon}_\alpha}.
\eqno{(3.4)}$$
In the same way we assign a ghost canonical pair to each $(a=1,2)$--pair of
the second--class constraint operators (2.8),

$$
T^a_\mu\quad\rightarrow\quad(C^{\prime\prime\mu},\CBPii_\mu),\quad
\mu=1,\ldots,m^{\prime\prime},
\eqno{(3.5)}$$

$$
\varepsilon(C^{\prime\prime\mu})=\varepsilon(\CBPii_\mu)=\varepsilon_\mu+1,
\eqno{(3.6)}$$

$$
\hbox{gh}^\prime(C^{\prime\prime\mu})=-\hbox{gh}^\prime(\CBPii_\mu)=0,\quad
\hbox{gh}^{\prime\prime}(C^{\prime\prime\mu})=
-\hbox{gh}^{\prime\prime}(\CBPii_\mu)=1
\eqno{(3.7)}$$

$$
(C^{\prime\prime\mu})^\dagger=C^{\prime\prime\mu},\quad
(\CBPii_\mu)^\dagger=-\CBPii_\mu(-1)^{\varepsilon_\mu}.
\eqno{(3.8)}$$
The equal-time nonzero supercommutators of the ghost operators introduced are

$$
(\imath\hbar)^{-1}[C^{\prime\alpha},\CBPi_\beta]=\delta^\alpha_\beta,\quad
(\imath\hbar)^{-1}[C^{\prime\prime\mu},\CBPii_\nu]=\delta^\mu_\nu.
\eqno{(3.9)}$$

Further, introduce the generating operators

$$
\Omega^a(q,p,C^\prime,\CBPi,C^{\prime\prime},\CBPii),\quad
\varepsilon(\Omega^a)=1,
\eqno{(3.10)}$$

$$
\hbox{gh}^\prime(\Omega^a)=0,\quad\hbox{gh}^{\prime\prime}(\Omega^a)=1,
\eqno{(3.11)}$$

$$
\Omega(q,p,C^\prime,\CBPi,C^{\prime\prime},\CBPii),\quad\varepsilon(\Omega)=1,
\eqno{(3.12)}$$

$$
\hbox{gh}^{\prime}(\Omega) =1,\quad\hbox{gh}^{\prime\prime}(\Omega)=0,
\eqno{(3.13)}$$

$$
K(q,p,C^\prime,\CBPi,C^{\prime\prime},\CBPii),\quad
\varepsilon(K)=0,
\eqno{(3.14)}$$

$$
\hbox{gh}^\prime(K)=2,\quad\hbox{gh}^{\prime\prime}(K)=-2,
\eqno{(3.15)}$$

$$
\CH(q,p,C^\prime,\CBPi,C^{\prime\prime},\CBPii),\quad
\varepsilon(\CH)=0,
\eqno{(3.16)}$$

$$
\hbox{gh}^\prime(\CH)=0,\quad\hbox{gh}^{\prime\prime}(\CH)=0,
\eqno{(3.17)}$$

$$
\Lambda(q,p,C^\prime,\CBPi,C^{\prime\prime},\CBPii),\quad
\varepsilon(\Lambda)=1,
\eqno{(3.18)}$$

$$
\quad\hbox{gh}^\prime(\Lambda)=1,\quad
\hbox{gh}^{\prime\prime}(\Lambda)=-2,
\eqno{(3.19)}$$
and subordinate them to the following generating equations:

$$
[\Omega^a,\Omega^b]=0,\quad(\Omega^a)^\dagger=\Omega^a,
\eqno{(3.20)}$$

$$
[\Omega^a,\Omega]=0,\quad(\Omega)^\dagger=\Omega,
\eqno{(3.21)}$$

$$
[\Omega,\Omega]=\varepsilon_{ab}(\imath\hbar)^{-1}[\Omega^b,[\Omega^a,K]],
\quad(K)^\dagger=K,
\eqno{(3.22)}$$

$$
[\Omega^a,\CH]=0,\quad(\CH)^\dagger=\CH,
\eqno{(3.23)}$$

$$
[\Omega,\CH]=\varepsilon_{ab}(\imath\hbar)^{-1}[\Omega^b,[\Omega^a,\Lambda]],
\quad(\Lambda)^\dagger=\Lambda.
\eqno{(3.24)}$$

Let us seek for a solution to these equations in the form of $C\CBP$--ordered
series expansion in ghost powers:

$$
\Omega^a=C^{\prime\prime\mu}T^a_\mu+{1\over2}(-1)^{\varepsilon_\nu}
C^{\prime\prime\nu} C^{\prime\prime\mu}U^{a\rho}_{\mu\nu}\CBPii_\rho
(-1)^{\varepsilon_\rho}+(-1)^{\tilde{\varepsilon}_\alpha}C^{\prime\alpha}
C^{\prime\prime\mu}\tilde{U}^{a\beta}_{\mu\alpha}\CBPi_\beta
(-1)^{\tilde{\varepsilon}_\beta}+\ldots,
\eqno{(3.25)}$$

$$
\Omega=C^{\prime\alpha}T_\alpha+{1\over2}(-1)^{\tilde{\varepsilon}_\beta}
C^{\prime\beta}C^{\prime\alpha}\tilde{U}^\gamma_{\alpha\beta}\CBPi_\gamma
(-1)^{\tilde{\varepsilon}_\gamma}+(-1)^{\tilde{\varepsilon}_\alpha}
C^{\prime\alpha}C^{\prime\prime\mu}U^\nu_{\mu\alpha}\CBPii_\nu
(-1)^{\varepsilon_\nu}+\ldots,
\eqno{(3.26)}$$

$$
K={1\over2}(-1)^{\tilde{\varepsilon}_\beta}C^{\prime\beta}C^{\prime\alpha}
W^{\mu\nu}_{\alpha\beta}\CBPii_\nu\CBPii_\mu(-1)^{\varepsilon_\nu}+\ldots,
\eqno{(3.27)}$$

$$
\CH=H-C^{\prime\prime\mu}V^\nu_\mu\CBPii_\nu(-1)^{\varepsilon_\nu}-
C^{\prime\alpha}\tilde{V}^\beta_\alpha\CBPi_\beta
(-1)^{\tilde{\varepsilon}_\beta}+\ldots,
\eqno{(3.28)}$$

$$
\Lambda={1\over2}C^{\prime\alpha}W^{\mu\nu}_\alpha\CBPii_\nu\CBPii_\mu
(-1)^{\varepsilon_\nu}+\ldots.
\eqno{(3.29)}$$

Of course, we have chosen the $C\CBP$--ordering only for the sake of
convenience
of the general analysis. Depending on a particular representation of
constraints
some other choice of ghost ordering may appear to be more relevant, such as
the Weyl-- or Wick--ordering in field--theory case.

By inserting the expansions (3.25) -- (3.29) into the left generating equations
in (3.20) -- (3.24), one obtains to the second order in ghosts just the
constraint involution relations (2.11) -- (2.15), whereas to higher orders
in ghosts we obtain all the higher structure relations\footnote{In particular,
the relation (2.25) is generated by the left equation (3.20) to the
$C^\prime(C^{\prime\prime})^2\CBPi$--order, whereas the
corresponding contribution to $\Omega^a$ is of the form (see also eq. (A.10) of
the Appendix)
$$
{1\over2}(-1)^{(\varepsilon_\nu+\varepsilon_\mu\tilde{\varepsilon}_\alpha)}
C^{\prime\alpha}C^{\prime\prime\nu}C^{\prime\prime\mu}
\tilde{U}^{a\beta\rho}_{\mu\nu\alpha}\CBPii_\rho\CBPi_\beta
(-1)^{\varepsilon_\rho}.$$} of the gauge algebra initiated by the given
operators (2.5), (2.6), (2.8). On the other hand, the right equations in
(3.20) -- (3.24) determine the properties of the constraints and higher
structure coefficients with respect to the Hermitian conjugation. Thus the
equations (3.20) -- (3.24) describe the gauge algebra generating mechanism
comprehensively.

The following Existence Theorem holds for the proposed generating equations
(3.20) -- (3.24): if the constraint involution relations (2.11) -- (2.5) are
satisfied together with the conditions (2.20), (2.25) and the ones requiring
for the  first-class  constraints  $T_\alpha$ to be irreducible in the
above formulated sense, then there also exist all the higher structure
coefficients in the expansions (3.25) -- (3.29) and, thus, there exists a
formal solution of the algebra generating equations.  Besides,  it can
be shown that all the Hermiticity properties in (3.20) -- (3.24) can also
be satisfied by the solution obtained.

The algebra generating equations (3.20) -- (3.24) admit the following group of
automorphisms:

$$
A=A_1\cdot A_2\cdot A_3
\eqno{(3.30)}$$
where $A_1$ is the standard unitary group

$$
\Omega^a\quad\rightarrow\quad U^{-1}\Omega^aU,
\eqno{(3.31)}$$

$$
\Omega\quad\rightarrow\quad U^{-1}\Omega U,\quad K\quad\rightarrow\quad
U^{-1}KU,
\eqno{(3.32)}$$

$$
\CH\quad\rightarrow\quad U^{-1}\CH U,\quad\Lambda\quad\rightarrow\quad
U^{-1}\Lambda U,
\eqno{(3.33)}$$

$A_2=GL(2,R)$ is the group of c--numerical nondegenerate linear
transformations
$$
\Omega^a\quad\rightarrow S^a_b\Omega^b,\quad\Omega\quad\rightarrow\quad\Omega,
\quad\CH\quad\rightarrow\quad\CH,
\eqno{(3.34)}$$

$$
K\quad\rightarrow\quad\lambda^{-1}K,\quad\Lambda\quad\rightarrow\quad
\lambda^{-1}\Lambda,\quad\lambda\equiv\hbox{det}(S^a_b),
\eqno{(3.35)}$$
$A_3$ is the group of exact shifts

$$
\Omega^a\quad\rightarrow\quad\Omega^a,
\eqno{(3.36)}$$

$$
\Omega\quad\rightarrow\quad\Omega+\varepsilon_{ab}(\imath\hbar)^{-2}[\Omega^b,
[\Omega^a,\Xi]],
\eqno{(3.37)}$$

$$
K\quad\rightarrow\quad K+2(\imath\hbar)^{-1}[\Omega,\Xi]+\varepsilon_{ab}
(\imath\hbar)^{-3}[[\Omega^b,\Xi],[\Omega^a,\Xi]]+(\imath\hbar)^{-1}[\Omega^a,
X_a],
\eqno{(3.38)}$$

$$
\CH\quad\rightarrow\quad\CH+(\imath\hbar)^{-1}[\Omega,\Psi]+\varepsilon_{ab}
(\imath\hbar)^{-1}[\Omega^b,[\Omega^a,\Phi]],\quad[\Omega^a,\Psi]=0,
\eqno{(3.39)}$$

$$\begin{array}{c}
\Lambda\quad\rightarrow\quad\Lambda+(\imath\hbar)^{-1}[\Xi,\CH]+
(\imath\hbar)^{-1}[\Omega,\Phi]+{1\over2}(\imath\hbar)^{-1}[K,\Psi]+\\[9pt]
+(\imath\hbar)^{-2}[\Xi,[\Omega,\Psi]]+\varepsilon_{ab}(\imath\hbar)^{-3}
[[\Xi,\Omega^b],[\Omega^a,\Phi]]+(\imath\hbar)^{-1}[\Omega^a,Y_a].
\end{array}\eqno{(3.40)}$$

Under the premises of the Existence Theorem the group of automorphisms
(3.30) is  the maximal possible one and,  thus,  describes the natural
arbitrariness of  a  solution  to  the  algebra  generating  equations
(3.20) -- (3.24) comprehensively.

The exact  shift  transformations (3.37) -- (3.40) enable one to make the
new operators $\bar{K}$ and $\bar{\Lambda}$ vanish. Then one can apply the
ghost--dependent canonical  transformations (3.31) -- (3.33) to make the
generating operators take the Abelian form

$$
\Omega^a_{abelian}=C^{\prime\prime\mu}t_\mu,\quad\Omega_{abelian}=
C^{\prime\alpha}t_\alpha,
\eqno{(3.41)}$$

$$
[t^{\{a}_\mu,t^{b\}}_\nu]=0,\quad[t^a_\mu,t_\alpha]=0,\quad[t_a,t_\beta]=0,
\eqno{(3.42)}$$

$$
[\CH_{abelian},t^a_\mu]=0,\quad[\CH_{abelian},t_\alpha]=0.
\eqno{(3.43)}$$

\section{Unitarizing Hamiltonian}

Introduce now the following new canonical variable operators which are the
antighosts:

$$
(\CP^{\prime\alpha},\bar{C}^\prime_\alpha),\quad\alpha=1,\dots,m^\prime
\eqno{(4.1)}$$

$$
\varepsilon(\CP^{\prime\alpha})=\varepsilon(\bar{C}^\prime_\alpha)=
\tilde{\varepsilon}_\alpha+1,
\eqno{(4.2)}$$

$$
\hbox{gh}^\prime(\CP^{\prime\alpha})=-\hbox{gh}^\prime(\bar{C}^\prime_\alpha)=
1,\quad\hbox{gh}^{\prime\prime}(\CP^{\prime\alpha})=
-\hbox{gh}^{\prime\prime}(\bar{C}^\prime_\alpha)=0,
\eqno{(4.3)}$$

$$
(\CP^{\prime\alpha})^\dagger=\CP^{\prime\alpha},\quad
(\bar{C}^\prime_\alpha)^\dagger=-\bar{C}^\prime_\alpha
(-1)^{\tilde{\varepsilon}_\alpha},
\eqno{(4.4)}$$

$$
(\CP^{\prime\prime\mu},\bar{C}^{\prime\prime}_\mu),\quad\mu=1,\dots,m^
{\prime\prime},
\eqno{(4.5)}$$

$$
\varepsilon(\CP^{\prime\prime\mu})=
\varepsilon(\bar{C}^{\prime\prime}_\mu)=\varepsilon_\mu+1,
\eqno{(4.6)}$$

$$
\hbox{gh}^\prime(\CP^{\prime\prime\mu})=
-\hbox{gh}^\prime(\bar{C}^{\prime\prime}_\mu)= 0,\quad
\hbox{gh}^{\prime\prime}(\CP^{\prime\prime^\mu})=
-\hbox{gh}^{\prime\prime}(\bar{C}^{\prime\prime}_\mu)=1,
\eqno{(4.7)}$$

$$
(\CP^{\prime\prime\mu})^\dagger=\CP^{\prime\prime\mu},\quad
(\bar{C}^{\prime\prime}_\mu)^\dagger=-\bar{C}^{\prime\prime}_\mu
(-1)^{\varepsilon_\mu},
\eqno{(4.8)}$$
and dynamically-active Lagrange multipliers:

$$
(\lambda^\alpha,\pi_\alpha),\quad\alpha=1,\dots,m^\prime,
\eqno{(4.9)}$$

$$
\varepsilon(\lambda_\alpha)=
\varepsilon(\pi_\alpha)=\tilde{\varepsilon}_\alpha,
\eqno{(4.10)}$$

$$
\hbox{gh}^\prime(\lambda^\alpha)=-\hbox{gh}^\prime(\pi_\alpha)=0,\quad
\hbox{gh}^{\prime\prime}(\lambda^\alpha)=-\hbox{gh}^{\prime\prime}
(\pi_\alpha)=0,
\eqno{(4.11)}$$

$$
(\lambda^\alpha)^\dagger=\lambda^\alpha(-1)^{\tilde{\varepsilon}_\alpha},\quad
(\pi_\alpha)^\dagger=\pi_\alpha,
\eqno{(4.12)}$$

$$
(\lambda^a_\mu),\quad a=1,2,\quad\mu=1,\dots,m^{\prime\prime},
\eqno{(4.13)}$$

$$
\varepsilon(\lambda^a_\mu)= \varepsilon_\mu,
\eqno{(4.14)}$$

$$
\quad\hbox{gh}^\prime(\lambda^a_\mu)=
\hbox{gh}^{\prime\prime}(\lambda^a_\mu)=0.
\eqno{(4.15)}$$

$$
(\lambda^a_\mu)^\dagger=\lambda^a_\mu.
\eqno{(4.16)}$$
The equal--time nonzero supercommutators of the new operators introduced are

$$
(\imath\hbar)^{-1}[\CP^{\prime\alpha},\bar{C}^\prime_\beta]=\delta^\alpha_\beta,
\quad(\imath\hbar)^{-1}[\CP^{\prime\prime\mu},\bar{C}^{\prime\prime}_\nu]=
\delta^\mu_\nu,
\eqno{(4.17)}$$

$$
(\imath\hbar)^{-1}[\lambda^\alpha,\pi_\beta]=\delta^\alpha_\beta,\quad
(\imath\hbar)^{-1}[\lambda^a_\mu,\lambda^b_\nu]=\varepsilon^{ab}d_{\mu\nu},
\eqno{(4.18)}$$
where a constant matrix $d_{\mu\nu}$ is supposed to be invertible and
possesses the following symmetry properties

$$
d_{\nu\mu}=d_{\mu\nu}(-1)^{\varepsilon_\mu\varepsilon_\nu},\qquad
d^*_{\nu\mu}=d_{\mu\nu}.
\eqno{(4.19)}$$

Let us extend the generating operators (3.10), (3.12) by including the
phase variable operators (4.1), (4.5), (4.9), (4.14) via the formulae

$$
Q=\Omega+\CP^{\prime\alpha}\pi_\alpha,
\eqno{(4.20)}$$

$$
Q^a=\Omega^a+\CP^{\prime\prime\mu}\lambda^a_\mu,\quad a=1,2,
\eqno{(4.21)}$$
so that

$$
\varepsilon(Q)=1,\quad\hbox{gh}^\prime(Q)=1,\quad\hbox{gh}^{\prime\prime}(Q)=0,
\eqno{(4.22)}$$

$$
\varepsilon(Q^a)=1,\quad\hbox{gh}^\prime(Q^a)=0,\quad
\hbox{gh}^{\prime\prime}(Q^a)=1.
\eqno{(4.23)}$$

The extended   operators   $Q$,   $Q^a$  satisfy  the  same  equations
(3.20) -- (3.24) as their minimal--sector counterparts $\Omega$, $\Omega^a$
do.

The complete Unitarizing Hamiltonian of the theory reads

$$
H_{complete}=\CH+(\imath\hbar)^{-1}[Q,F]+\varepsilon_{ab}(\imath\hbar)^{-2}
[Q^b,[Q^a,B]],\quad[Q^a,F]=0,
\eqno{(4.24)}$$
where

$$
\varepsilon(F)=1,\quad\hbox{gh}^\prime(F)=-1,\quad\hbox{gh}^{\prime\prime}(F)=0,
\eqno{(4.25)}$$

$$
\varepsilon(B)=0,\quad\hbox{gh}^\prime(B)=0,\quad\hbox{gh}^{\prime\prime}(B)
=-2.
\eqno{(4.26)}$$

$$
(F)^\dagger=-F,\quad (B)^\dagger=-B.
\eqno{(4.27)}$$
The gauge--fixing operators $F$ and $B$ may depend on the total set of phase
variables of the extended phase space. In the simplest case these gauge
operators can be chosen in the form

$$
F=\lambda^\alpha\CBPi_\alpha+(\chi^\alpha+C^{\prime\prime\mu}V^{\nu\alpha}_\mu
\CBPii_\nu(-1)^{\varepsilon_\nu+\varepsilon_\alpha})\bar{C}^\prime_\alpha+
\ldots, \quad(\imath\hbar)^{-1}[T^a_\mu,\chi^\alpha]=V^{\nu\alpha}_\mu T^a_\nu,
\eqno{(4.28)}$$

$$
B=\CBPii_\mu\bar{C}^{\prime\prime}_\nu d^{\nu\mu},\quad d_{\mu\nu}d^{\nu\rho}=
\delta^\rho_\mu.
\eqno{(4.29)}$$

Further, let us introduce the ghost number operators

$$
G^\prime={1\over2}(C^{\prime\alpha}\CBPi_\alpha(-1)^{\tilde{\varepsilon}_\alpha}
-\CBPi_\alpha C^{\prime\alpha})+{1\over2}(\CP^{\prime\alpha}
\bar{C}^\prime_\alpha(-1)^{\tilde{\varepsilon}_\alpha}-
\bar{C}^\prime_\alpha\CP^{\prime\alpha}), \eqno{(4.30)}$$

$$
G^{\prime\prime}={1\over2}(C^{\prime\prime\mu}\CBPii_\mu(-1)^{\varepsilon_\mu}
-\CBPii_\mu C^{\prime\prime\mu})+{1\over2}(\CP^{\prime\prime\mu}
\bar{C}^{\prime\prime}_\mu(-1)^{\varepsilon_\mu}-\bar{C}^{\prime\prime}_\mu
\CP^{\prime\prime\mu}).
\eqno{(4.31)}$$
Then we have

$$
(\imath\hbar)^{-1}[G^\prime,A]=\hbox{gh}^\prime(A)A,\quad
G^\prime|\Phi\rangle=\hbox{gh}^\prime(|\Phi\rangle)|\Phi\rangle,
\eqno{(4.32)}$$

$$
(\imath\hbar)^{-1}[G^{\prime\prime},A]=\hbox{gh}^{\prime\prime}(A)A,\quad
G^{\prime\prime}|\Phi\rangle=\hbox{gh}^{\prime\prime}(|\Phi\rangle)|\Phi\rangle.
\eqno{(4.33)}$$
The total ghost number operator is naturally defined as

$$
G=G^\prime+G^{\prime\prime}.
\eqno{(4.34)}$$

As a next step, let us define the physical operators and physical states.
An operator $\CO$ is called the physical one iff

$$
\hbox{gh}^\prime(\CO)=\hbox{gh}^{\prime\prime}(\CO)=0,
\eqno{(4.35)}$$

$$
[Q^a,\CO]=0,\quad[Q,\CO]=\varepsilon_{ab}(\imath\hbar)^{-1}
[Q^b,[Q^a,E]].
\eqno{(4.36)}$$
Of course,  the  Hamiltonian (4.24) is a physical operator just in the
sense  of this definition.

A state $|\Phi\rangle$ is called the physical one iff

$$
\hbox{gh}^\prime(|\Phi\rangle)=\hbox{gh}^{\prime\prime}(|\Phi\rangle)=0,
\eqno{(4.37)}$$

$$
Q^a|\Phi\rangle=0,\quad
Q|\Phi\rangle=\varepsilon_{ab}(\imath\hbar)^{-1}Q^bQ^a|E\rangle.
\eqno{(4.38)}$$
The physical matrix elements $\langle\Phi|\CO|\Phi_1\rangle$ depend neither on
the arbitrariness (3.30) in determining the generating operators
$\Omega^a$, $\Omega$, $K$, $\CH$, $\Lambda$, nor on the arbitrariness of
r.h.s. of eqs. (4.36), (4.38).

Let $\Gamma$ be the total set of phase variable operators of the extended
phase space, and let $\Gamma(t)$ satisfies the Heisenberg equations governed
by the  Unitarizing Hamiltonian (4.24).  Then the physical matrix elements
$\langle\Phi|\CO(\Gamma(t))|\Phi_1\rangle$ do not depend on a particular choice
of gauge--fixing  operators $F$ and $B$.

\section{Further Generalization and Geometric Interpretation}

It has been implied in the above considerations that the second--class
constraints themselves retain their algebraic properties to be the same
as they are in the pure second--class case. In particular, no first--class
constraints enter the split involution relations (1.1), (1.2) actually.

In this section we intend to generalize the set of constraint algebra
generating equations in order to make  it possible for the first--class
constraints contribute explicitly to the modified split involution relations.

The main idea can be explained as follows. Let the original second--class
constraints $T^a_\mu$ are allowed to contain the first--class admixture.
Let us suppose that the corresponding admixture to the generating operators
$\Omega^a$ is representable in the form

$$
(\imath\hbar)^{-1}[A^a,\Omega],
\eqno{(5.1)}$$
where new ghost--dependent operators $A^a$ are introduced,

$$
\varepsilon(A^a)=0,\quad\hbox{gh}^\prime(A^a)=-1,\quad\hbox{gh}^{\prime\prime}
(A^a)=1,\quad(A^a)^\dagger=A^a,
\eqno{(5.2)}$$
and $\Omega$ is the first--class generating operator to be determined
selfconsistently.

It is quite natural to require for the pure second--class generating
operators

$$
\Omega^a-(\imath\hbar)^{-1}[A^a,\Omega]
\eqno{(5.3)}$$
to satisfy the equations similar to the above--given ones (3.20), (3.21):

$$
[\Omega^a-(\imath\hbar)^{-1}[A^a,\Omega],
\Omega^b-(\imath\hbar)^{-1}[A^b,\Omega]]=0,
\eqno{(5.4)}$$

$$
[\Omega^a-(\imath\hbar)^{-1}[A^a,\Omega],\Omega]=0,
\eqno{(5.5)}$$

Besides, we have to subordinate the first--class generating operator $\Omega$
to
the equation similar the one (3.22):

$$
[\Omega,\Omega]=\varepsilon_{ab}(\imath\hbar)^{-1}
[\Omega^b-(\imath\hbar)^{-1}[A^b,\Omega],[\Omega^a-(\imath\hbar)^{-1}[A^a,
\Omega],K]].
\eqno{(5.6)}$$
In the same way we formulate the equations similar to the above--given ones
(3.23), (3.24):

$$
[\Omega^b-(\imath\hbar)^{-1}[A^b,\Omega],\CH]=0,
\eqno{(5.7)}$$

$$
[\Omega,\CH]=\varepsilon_{ab}(\imath\hbar)^{-1}
[\Omega^b-(\imath\hbar)^{-1}[A^b,\Omega],[\Omega^a-(\imath\hbar)^{-1}[A^a,
\Omega],\Lambda]].
\eqno{(5.8)}$$

The generating operators $\Omega^a$, $\Omega$, $K$, $\CH$, $\Lambda$ are
searched in the form of the corresponding series expansions (3.25) -- (3.29),
whereas the new operators $A^a$ are expanded in ghost powers as

$$\begin{array}{c}
A^a=C^{\prime\prime\mu}\tilde{X}^{a\beta}_\mu\CBPi_\beta
(-1)^{\tilde{\varepsilon}_\beta}+{1\over2}(-1)^{\tilde{\varepsilon}_\alpha}
C^{\prime\alpha}C^{\prime\prime\mu}\tilde{\tilde{X}}
{}^{a\beta\gamma}_{\mu\alpha}\CBPi\gamma\CBPi_\beta
(-1)^{\tilde{\varepsilon}_\gamma}+\\[9pt]
+{1\over2}(-1)^{\varepsilon_\nu}C^{\prime\prime\nu}C^{\prime\prime\mu}
\tilde{X}^{a\alpha\rho}_{\mu\nu}\CBPii_\rho\CBPi_\alpha
(-1)^{\varepsilon_\rho}+\ldots \,\,.
\end{array}\eqno{(5.9)}$$

Here we refrain from considering in details the explicit form of a constraint
algebra generated by eqs. (5.4) -- (5.8) to the lowest order in ghosts. The
only
comment to be given here concerns the modified cross--sector relations. Instead
of (2.12) we have:

$$\begin{array}{c}
(\imath\hbar)^{-1}[T^a_\mu,T_\alpha]=\tilde{U}^{a\beta}_{\mu\alpha}T_\beta+
U^\nu_{\mu\alpha}T^a_\nu+\\[9pt]
+{1\over2}(\delta^\gamma_\alpha\tilde{X}^{a\beta}_\mu-\delta^\beta_\alpha
\tilde{X}^{a\gamma}_\mu
(-1)^{\tilde{\varepsilon}_\beta\tilde{\varepsilon}_\gamma}+\imath\hbar
\tilde{\tilde{X}}{}^{a\gamma\beta}_{\mu\alpha})((\imath\hbar)^{-1}[T_\beta,
T_\gamma]-\tilde{U}^\delta_{\beta\gamma}T_\delta).
\end{array}\eqno{(5.10)}$$
Being treated at the classical level, these  relations
determine the quantities $\tilde{X}^{a\beta}_\mu$ to serve as coefficients of
a
linear  dependence  between   the  cross--sector  supercommutators
$\{T^a_\mu,
T_\alpha\}$ and the pure first--class--sector ones $\{T_\alpha, T_\beta\}$.

The following Existence Theorem apparently holds for the generating equations
(5.4) -- (5.8): if these equations are satisfied to the lowest order in ghosts
and, besides, the equations (5.4) themselves are satisfied to the
$C^\prime(C^{\prime\prime})^2\CBPi$--order, then there exists a formal solution
for the generating operators $\Omega^a$, $\Omega$, $K$, $\CH$, $\Lambda$ to all
orders in ghosts.

It is an interesting circumstance that l.h.s. of eqs. (5.4) -- (5.8)
possess the structure of a natural first--class counterpart of the well--known
Dirac's bracket, being eq.(5.5) represented in the equivalent form

$$
(\imath\hbar)^{-1}[A^a,{1\over2}[\Omega,\Omega]]=[\Omega^a,\Omega]
\eqno{(5.11)}$$
to determine the "Lagrange multiplier" operators $A^a$ to an admissible extent.
It is just the form that generalizes in the most natural way the lowest--order
relations (5.10).

Now, let us consider an interesting geometric extension to the set of eqs.
(5.4) -- (5.8). First of all, introduce a pair of real Bosonic ghost
parameters $\xi_a$,

$$
\varepsilon(\xi_a)=0,\quad\hbox{gh}^\prime(\xi_a)=1,\quad
\hbox{gh}^{\prime\prime}(\xi_a)=-1,\quad\xi^*_a=\xi_a.
\eqno{(5.12)}$$

Next, let us define the $\xi$--dependent generating operators

$$
\bar{\Omega},\quad\bar{K},\quad\bar{\CH},\quad\bar{\Lambda},\quad\bar{A^a}
\eqno{(5.13)}$$
to satisfy the equations

$$
[\bar{\Omega},\bar{\Omega}]=\varepsilon_{ab}(\imath\hbar)^{-1}
[\nabla^b\bar{\Omega},[\nabla^a\bar{\Omega},\bar{K}]],
\eqno{(5.14)}$$

$$
\nabla^a\nabla^b\bar{\Omega}=0,\quad\nabla^a\bar{K}=0,
\eqno{(5.15)}$$

$$
[\bar{\Omega},\bar{\CH}]=\varepsilon_{ab}(\imath\hbar)^{-1}[\nabla^b\bar{\Omega},
[\nabla^a\bar{\Omega},\bar{\Lambda}]],
\eqno{(5.16)}$$

$$
\nabla^a\bar{\CH}=0,\quad\nabla^a\bar{\Lambda}=0,
\eqno{(5.17)}$$
where

$$
\nabla^a\equiv\partial^a-(\imath\hbar)^{-1}\hbox{ad}\bar{A}^a,\quad
\partial^a\equiv{\partial\over\partial\xi_a}
\eqno{(5.18)}$$
stand for the covariant $\xi$--derivative components, so that for arbitrary
$E(\xi)$ we have

$$
\nabla^aE=\partial^a E-(\imath\hbar)^{-1}[\bar{A}^a,E].
\eqno{(5.19)}$$
We suppose the connection $\bar{A}^a$ to be flat:

$$
\partial^{[a}\bar{A}^{b]}-(\imath\hbar)^{-1}[\bar{A}^a,\bar{A}^b]=0.
\eqno{(5.20)}$$
It follows immediately from eqs. (5.14) -- (5.17), (5.20) that

$$
\nabla^a[\bar{\Omega},\bar{\Omega}]\equiv2[\nabla^a\bar{\Omega},\bar{\Omega}]=0,
\eqno{(5.21)}$$

$$
[\nabla^a\bar{\Omega},\nabla^b\bar{\Omega}]=0,
\eqno{(5.22)}$$

$$
[\nabla^a\bar{\Omega},\bar{\CH}]=0.
\eqno{(5.23)}$$
If one denotes:

$$
\bar{\Omega}\big|_{\xi=0}\equiv\Omega,\quad
\partial^a\bar{\Omega}\big|_{\xi=0}\equiv\Omega^a,
\eqno{(5.24)}$$

$$
\bar{A}^a\big|_{\xi=0}\equiv A^a,\quad\bar{K}\big|_{\xi=0}\equiv K,
\eqno{(5.25)}$$

$$
\bar{\CH}\big|_{\xi=0}\equiv\CH,\quad\bar{\Lambda}\big|_{\xi=0}\equiv\Lambda,
\eqno{(5.26)}$$
the equations (5.14), (5.16), (5.21) -- (5.23), being taken at $\xi=0$, just
reproduce the set of eqs. (5.4) -- (5.8).

Due to the flatness condition (5.20) there exists a $\xi$--dependent canonical
transformation that results for the connection components $\bar{A}^a$ in their
vanishing. Thus one returns naturally to the case considered in previous
Sections.

Further, let us extend the operator $\bar{\Omega}$ via the formula

$$
\bar{Q}\equiv\bar{\Omega}+\CP^{\prime\alpha}\pi_\alpha+\CP^{\prime\prime\mu}
\lambda^a_\mu\xi_a.
\eqno{(5.27)}$$
Then the $\xi$--dependent Unitarizing Hamiltonian reads

$$
\bar{H}_{complete}\equiv\bar{\CH}+(\imath\hbar)^{-1}[\bar{Q},\bar{F}]+
\varepsilon_{ab}(\imath\hbar)^{-2}[\nabla^b\bar{Q},[\nabla^a\bar{Q},\bar{B}]],
\eqno{(5.28)}$$
where $\xi$--dependent gauge--fixing operators $\bar{F}$, $\bar{B}$ should
satisfy the conditions

$$
\nabla^a\bar{F}=0\quad[\bar{F},\nabla^a\bar{Q}]=0,\quad\nabla^a\bar{B}=0.
\eqno{(5.29)}$$

Finally, let us define the $\xi$--dependent physical operators and states. An
operator $\bar{\CO}$ is called the physical one iff:

$$
\hbox{gh}^\prime(\bar{\CO})=\hbox{gh}^{\prime\prime}(\bar{\CO})=0,
\eqno{(5.30)}$$

$$
[\bar{Q},\bar{\CO}]=\varepsilon_{ab}(\imath\hbar)^{-1}[\nabla^b\bar{Q},
[\nabla^a\bar{Q},\bar{E}]],
\eqno{(5.31)}$$

$$
\nabla^a\bar{\CO}=0,\quad\nabla^a\bar{E}=0.
\eqno{(5.32)}$$

A state $|\bar{\Phi}\rangle$ is called the physical one iff

$$
\hbox{gh}^\prime(|\bar{\Phi}\rangle)=
\hbox{gh}^{\prime\prime}(|\bar{\Phi}\rangle)=0,
\eqno{(5.33)}$$

$$
\bar{Q}|\bar{\Phi}\rangle=\varepsilon_{ab}(\imath\hbar)^{-1}(\nabla^b\bar{Q})
(\nabla^a\bar{Q})|\bar{E}\rangle,
\eqno{(5.34)}$$

$$
\nabla^a|\bar{\Phi}\rangle=0,\quad\nabla^a|\bar{E}\rangle=0,
\eqno{(5.35)}$$
where the covariant derivative operators $\nabla^a$ are applied to arbitrary
state $|\ldots\rangle$ via the formula: $\nabla^a|\ldots\rangle\equiv
(\partial^a-(\imath\hbar)^{-1}A^a)|\ldots\rangle$.

By construction, the physical matrix elements are $\xi$--independent:

$$
\langle\bar{\Phi}|\bar{\CO}|\bar{\Phi}_1\rangle=
\langle\Phi|\CO|\Phi_1\rangle
\eqno{(5.36)}$$
where unbared operators and states in r.h.s. coincide with the corresponding
bared ones taken at $\xi=0$. (Of course, one can choose another fixed point
$\xi_0$ instead of $\xi=0$.)

The physical matrix elements (5.36) are also independent of a particular
choice of operators $F$, $B$, $E$ and states $|E\rangle$ entering eqs.
(5.28) -- (5.35) taken at $\xi=0$.

\section{Conclusion}

So, we have extended the split involution formalism to cover the case of the
presence of irreducible first--class constraints. Thereby the miraculous
supersymmetry yielded by the split involution relations is coupled to the
actual gauge symmetry initiated by the original first--class constraints.

The most characteristic feature of the formalism proposed is the appearance of
the new equivalence criterion explicitly--quadratic in second--class
constraints
that is a natural counterpart to the Dirac's weak equality concept as applied
to
the first--class quantities.

It is quite evident from this viewpoint that all the double--supercommuta\-tor
contributions in (3.22), (3.24), (3.39), (4.24), (4.36) as well as
the quadra\-tic operator in r.h.s. of (4.38) are of the same origin.

All the  main results are extendable in a straightforward way to cover
the case of finite--stage reducibility of the first  and  second-class
constraints included.

{\bf Acknowledgement.} The work is supported, in part, by the International
Science Foundation under the Grant number M2I000. The participation of
I.A.B. and S.L.L. is also supported by the European Community Commission
under the contract INTAS--93--633 and INTAS--93--2058, respectively.

\section{Appendix. Quantum Rules of Dividing by Constraints}

In this Appendix we represent the general solution to the equation (2.21) --
(2.24).

First of all, let us introduce the following remarkable operators :

$$
W_n\equiv\sum_{m=0}^{n-1}\Omega_m(C,\CBP)\big|_{C\rightarrow-\imath\hbar
{\partial_r\over\partial\CBP}(-1)^{\varepsilon(\CBP)}},
\eqno{(A.1)}$$

$$
W^a_n\equiv\sum_{m=0}^{n-1}\Omega^a_m(C,\CBP)\big|_{C\rightarrow-\imath\hbar
{\partial_r\over\partial\CBP}(-1)^{\varepsilon(\CBP)}},
\eqno{(A.2)}$$
where $C\equiv(C^\prime,C^{\prime\prime})$, $\CBP\equiv(\CBPi,\CBPii)$ is a
condensed notation for ghost operators, and

$$
\Omega_m\quad\sim\quad(C)^{m+1}(\CBP)^m,\quad\quad
\Omega^a_m\quad\sim\quad(C)^{m+1}(\CBP)^m
\eqno{(A.3)}$$
are the corresponding homogeneous monomials entering the ghost power series
expansions to the generating operators $\Omega$, $\Omega^a$,

$$
\Omega=\sum_{m=0}^\infty\Omega_m,\quad\Omega^a=\sum_{m=0}^\infty\Omega^a_m.
\eqno{(A.4)}$$
In particular we have

$$
\Omega_0=C^{\prime\alpha}T_\alpha,
\eqno{(A.5)}$$

$$
\Omega_1={1\over2}(-1)^{\tilde{\varepsilon}_\beta}C^{\prime\beta}
C^{\prime\alpha}\tilde{U}^\gamma_{\alpha\beta}\CBPi_\gamma
(-1)^{\tilde{\varepsilon}_\gamma}+(-1)^{\tilde{\varepsilon}_\alpha}
C^{\prime\alpha}C^{\prime\prime\mu}U^\nu_{\mu\alpha}\CBPii_\nu
(-1)^{\varepsilon_\nu},
\eqno{(A.6)}$$

$$\begin{array}{c}
\Omega_2={1\over12}(-1)^{(\tilde{\varepsilon}_\beta+\tilde{\varepsilon}_\alpha
\tilde{\varepsilon}_\gamma)}C^{\prime\gamma}C^{\prime\beta}C^{\prime\alpha}
\tilde{\tilde{U}}{}^{\delta\lambda}_{\alpha\beta\gamma}\CBPi_\lambda\CBPi_\delta
(-1)^{\tilde{\varepsilon}_\lambda}+\\[9pt]
+{1\over2}(-1)^{(\tilde{\varepsilon}_\alpha+
\tilde{\varepsilon}_\beta\varepsilon_\mu)}C^{\prime\beta}C^{\prime\alpha}
C^{\prime\prime\mu}\tilde{U}^{\gamma\nu}_{\mu\alpha\beta}\CBPii_\nu
\CBPi_\gamma(-1)^{\varepsilon_\nu}+\\[9pt]
{1\over4}(-1)^{(\varepsilon_\nu+
\tilde{\varepsilon}_\alpha\varepsilon_\mu)}C^{\prime\alpha}C^{\prime\prime\nu}
C^{\prime\prime\mu}U^{\rho\sigma}_{\mu\nu\alpha}\CBPii_\sigma\CBPii_\rho
(-1)^{\varepsilon_\sigma},
\end{array}\eqno{(A.7)}$$

$$
\Omega^a_0=C^{\prime\prime\mu}T^a_\mu,
\eqno{(A.8)}$$

$$
\Omega^a_1={1\over2}(-1)^{\varepsilon_\nu}C^{\prime\prime\nu}
C^{\prime\prime\mu}U^{a\rho}_{\mu\nu}\CBPii_\rho(-1)^{\varepsilon_\rho}+
(-1)^{\tilde{\varepsilon}_\alpha}C^{\prime\alpha}C^{\prime\prime\mu}
\tilde{U}^{a\beta}_{\mu\alpha}\CBPi_\beta(-1)^{\tilde{\varepsilon}_\beta},
\eqno{(A.9)}$$

$$\begin{array}{c}
\Omega^a_2={1\over12}(-1)^{(\varepsilon_\nu+\varepsilon_\mu\varepsilon_\rho)}
C^{\prime\prime\rho}C^{\prime\prime\nu}C^{\prime\prime\mu}
U^{a\nu\sigma\tau}_{\mu\nu\rho}\CBPii_\tau\CBPii_\sigma
(-1)^{\varepsilon_\tau}+\\[9pt]
+{1\over2}(-1)^{(\varepsilon_\nu+\tilde{\varepsilon}_\alpha\varepsilon_\mu)}
C^{\prime\alpha}C^{\prime\prime\nu}C^{\prime\prime\mu}
\tilde{U}^{a\beta\rho}_{\mu\nu\alpha}\CBPii_\rho\CBPi_\beta
(-1)^{\varepsilon_\rho}+\\[9pt]
{1\over4}(-1)^{(\tilde{\varepsilon}_\alpha+
\tilde{\varepsilon}_\beta\varepsilon_\mu)}C^{\prime\beta}C^{\prime\alpha}
C^{\prime\prime\mu}\tilde{\tilde{U}}{}^{a\gamma\delta}_{\mu\alpha\beta}
\CBPi_\delta\CBPi_\gamma(-1)^{\tilde{\varepsilon}_\delta}.
\end{array}\eqno{(A.10)}$$

As applied from the right to arbitrary $C\CBP$--ordered polinomial of
the highest power $n$ in ghost momenta $\CBP$, the operators (A.1), (A.2)
possess the important formal properties

$$
W_nW^a_{n-1}+W^a_nW_{n-1}=0,\quad W^{\{a}_nW^{b\}}_{n-1}=0,
\eqno{(A.11)}$$

$$
W_nW_{n-1}={1\over2}\Delta_n\big|_{C\rightarrow-\imath\hbar
{\partial_r\over\partial\CBP}(-1)^{\varepsilon(\CBP)}},
\eqno{(A.12)}$$
where

$$
\Delta\equiv\varepsilon_{ab}(\imath\hbar)^{-1}[\Omega^b,[\Omega^a,K]]=
\sum_{n=2}^\infty\Delta_n,\quad\quad\Delta_n\quad\sim\quad(C)^n(\CBP)^{n-2},
\eqno{(A.13)}$$

$$
\Delta_2=\bar{W}_2W^b_2W^a_1\varepsilon_{ab},\quad\bar{W}_2=(\imath\hbar)^{-1}
K_2\big|_{C\rightarrow-\imath\hbar{\partial_r\over\partial\CBP}
(-1)^{\varepsilon(\CBP)}},
\eqno{(A.14)}$$
and $K_m$ is the $(\CBP)^m$--order in the expansion (3.27).

Now, let us consider the equation (2.21) to represent it in the form

$$
Z_1W^a_1+\tilde{Z}^a_1W_1=0,
\eqno{(A.15)}$$
where

$$
Z_1\equiv Z^\mu\CBPii_\mu(-1)^{\varepsilon_\mu},\quad\tilde{Z}^a_1\equiv
\tilde{Z}^{a\alpha}\CBPi_\alpha(-1)^{\tilde{\varepsilon}_\alpha}.
\eqno{(A.16)}$$
It can be shown that the general solution for $Z_1$, $Z^a_1$ is

$$
Z_1=(E_3W^b_3+\tilde{\tilde{E}}{}^b_2\bar{W}_2)W^a_2\varepsilon_{ab}+
\tilde{E}_2W_2,
\eqno{(A.17)}$$

$$
\tilde{Z}^a_1=\tilde{E}_2W^a_2+\tilde{\tilde{E}}{}^a_2W_2,
\eqno{(A.18)}$$
where

$$
E_3\quad\sim\quad(\CBPii)^3,\quad\quad\tilde{E}_2\quad\sim\quad\CBPii\CBPi,
\quad\quad\tilde{\tilde{E}}{}^a_2\quad\sim\quad(\CBPi)^2
\eqno{(A.19)}$$
are arbitrary operators.

By eliminating ghost operators from the representations (A.17), (A.18), one
decodes the general solution for $Z^\mu$, $\tilde{Z}^{a\alpha}$ in the form

$$
Z^\mu=\tilde{E}^{\alpha\nu}\Pi^\mu_{\nu\alpha}+(E^{\tau\sigma\rho}
\Pi^{\xi\nu a}_{\rho\sigma\tau}+\imath\hbar\tilde{\tilde{E}}{}^{a\beta\alpha}
W^{\xi\nu}_{\alpha\beta})\Pi^{\mu b}_{\nu\xi}\varepsilon_{ba},
\eqno{(A.20)}$$

$$
\tilde{Z}^{a\alpha}=\tilde{E}^{\beta\mu}\tilde{\Pi}^{\alpha a}_{\mu\beta}+
\tilde{\tilde{E}}{}^{a\gamma\beta}\tilde{\Pi}^\alpha_{\beta\gamma},
\eqno{(A.21)}$$
where

$$
\Pi^\nu_{\mu\alpha}\equiv-T_\alpha\delta^\nu_\mu
(-1)^{\tilde{\varepsilon}_\alpha\varepsilon_\mu}-\imath\hbar U^\nu_{\mu\alpha},
\eqno{(A.22)}$$

$$
\Pi^{\sigma\tau a}_{\mu\nu\rho}\equiv[(\delta^\sigma_\mu
\bar{\Pi}^{\tau a}_{\nu\rho}-\delta^\tau_\mu\bar{\Pi}^{\sigma a}_{\nu\rho}
(-1)^{\varepsilon_\sigma\varepsilon_\tau})(-1)^{\varepsilon_\mu\varepsilon_\nu}+
\hbox{cycle}(\rho,\mu,\nu)]+(\imath\hbar)^2U^{a\sigma\tau}_{\mu\nu\rho},
\eqno{(A.23)}$$

$$
\bar{\Pi}^{\rho a}_{\mu\nu}\equiv{1\over2}(T^a_\mu\delta^\rho_\nu-
T^a_\nu\delta^\rho_\mu(-1)^{\varepsilon_\mu\varepsilon_\nu})-
\imath\hbar U^{a\rho}_{\mu\nu},
\eqno{(A.24)}$$

$$
\Pi^{\rho a}_{\mu\nu}\equiv T^a_\mu\delta^\rho_\nu-
T^a_\nu\delta^\rho_\mu(-1)^{\varepsilon_\mu\varepsilon_\nu}-
\imath\hbar U^{a\rho}_{\mu\nu},
\eqno{(A.25)}$$

$$
\tilde{\Pi}^{\beta a}_{\mu\alpha}\equiv
T^a_\mu\delta^\beta_\alpha-\imath\hbar\tilde{U}^{a\beta}_{\mu\alpha},
\eqno{(A.26)}$$

$$
\tilde{\Pi}^\gamma_{\alpha\beta}\equiv T_\alpha\delta^\gamma_\beta-
T_\beta\delta^\gamma_\alpha
(-1)^{\tilde{\varepsilon}_\alpha\tilde{\varepsilon}_\beta}-
\imath\hbar\tilde{U}^\gamma_{\alpha\beta}.
\eqno{(A.27)}$$

Next, let us consider the equation (2.22) to represent it in the form

$$
Z^{\{a}_1W^{b\}}_1+\tilde{Z}^{ab}_1W_1=0,\quad\tilde{Z}^{[ab]}=0,
\eqno{(A.28)}$$
where

$$
Z^a_1\equiv Z^{a\mu}\CBPii_\mu(-1)^{\varepsilon_\mu},\quad\tilde{Z}^{ab}_1
\equiv\tilde{Z}^{ab\alpha}\CBPi_\alpha(-1)^{\tilde{\varepsilon}_\alpha}.
\eqno{(A.29)}$$

The general solution for $Z^a_1$, $\tilde{Z}^{ab}_1$ is given by the formulae

$$
Z^a_1=E_2W^a_2+\tilde{E}^a_2W_2+{1\over2}\tilde{\tilde{E}}{}^{ac}_2\bar{W}_2
W^b_2\varepsilon_{bc},
\eqno{(A.30)}$$

$$
\tilde{Z}^{ab}_1=\tilde{E}^{\{a}_2W^{b\}}_2+\tilde{\tilde{E}}{}^{ab}_2W_2,\quad
\tilde{\tilde{E}}{}^{[ab]}_2=0,
\eqno{(A.31)}$$
where

$$
E_2\quad\sim\quad(\CBPii)^2,\quad\quad\tilde{E}^a_2\quad\sim\quad\CBPii\CBPi,
\quad\quad\tilde{\tilde{E}}{}^{ab}_2\quad\sim\quad(\CBPi)^2
\eqno{(A.32)}$$
are arbitrary operators.

By decoding the representations (A.30), (A.31) one obtains the general
solution for $Z^{a\mu}$, $\tilde{Z}^{ab\alpha}$ :

$$
Z^{a\mu}=(E^{\rho\nu}\delta^a_c+{1\over2}\imath\hbar
\tilde{\tilde{E}}{}^{ab\beta\alpha}W^{\rho\nu}_{\alpha\beta}\varepsilon_{cb})
\Pi^{\mu c}_{\nu\rho}+\tilde{E}^{a\alpha\nu}\Pi^\mu_{\nu\alpha},
\eqno{(A.33)}$$

$$
\tilde{Z}^{ab\alpha}=\tilde{E}^{\{a\beta\mu}
\tilde{\Pi}^{\alpha b\}}_{\mu\beta}+\tilde{\tilde{E}}{}^{ab\gamma\beta}
\tilde{\Pi}^\alpha_{\beta\gamma}.
\eqno{(A.34)}$$

Further, let us represent the equation (2.23) in the form

$$
Z^{ab}_1W^c_1+\hbox{cycle}(a,b,c)=0,\quad Z^{[ab]}_1=0,
\eqno{(A.35)}$$
where

$$
Z^{ab}_1\equiv Z^{ab\mu}\CBPii_\mu(-1)^{\varepsilon_\mu}.
\eqno{(A.36)}$$

The general solution is given by the formula

$$
Z^{ab}_1=E^{\{a}_2W^{b\}}_2
\eqno{(A.37)}$$
where

$$
E^a_2\quad\sim\quad(\CBPii)^2
\eqno{(A.38)}$$
are arbitrary operators.

It follows from (A.37) that the general solution for $Z^{ab\mu}$ is of the
form

$$
Z^{ab\mu}=E^{\{a\rho\nu}\Pi^{\mu b\}}_{\rho\nu}.
\eqno{(A.39)}$$

Finally, let us turn to the equation (2.24) as represented in the form

$$
Z_2W^b_2W^a_1\varepsilon_{ab}+\tilde{Z}_1W_1=0,
\eqno{(A.40)}$$
where

$$
Z_2\equiv-{1\over2}Z^{\mu\nu}\CBPii_\nu\CBPii_\mu(-1)^{\varepsilon_\nu},\quad
\tilde{Z}_1\equiv\tilde{Z}^\alpha\CBPi_\alpha(-1)^{\tilde{\varepsilon}_\alpha}
\eqno{(A.41)}$$
The general solution is given by the formulae

$$
Z_2=-\tilde{E}_3W_3+E^b_3W^a_3\varepsilon_{ab}-
{1\over2}\tilde{\tilde{E}}_2\bar{W}_2,
\eqno{(A.42)}$$

$$
Z_1=\tilde{\tilde{E}}_2W_2+\tilde{E}_3W^b_3W^a_2\varepsilon_{ab},
\eqno{(A.43)}$$
where

$$
E^a_3\quad\sim\quad(\CBPii)^3,\quad\quad
\tilde{E}_3\quad\sim\quad(\CBPii)^2\CBPi,\quad\quad
\tilde{\tilde{E}}_2\quad\sim\quad(\CBPi)^2
\eqno{(A.44)}$$
are arbitrary operators.

By decoding the representations (A.42), (A.43), one obtains the general
solution for $Z^{\mu\nu}$, $\tilde{Z}^\alpha$ in the form

$$
Z^{\mu\nu}=\tilde{E}^{\alpha\sigma\rho}\Pi^{\mu\nu}_{\rho\sigma\alpha}+
E^{b\tau\sigma\rho}\Pi^{\mu\nu a}_{\rho\sigma\tau}\varepsilon_{ab}-
\imath\hbar\tilde{\tilde{E}}{}^{\beta\alpha}W^{\mu\nu}_{\alpha\beta},
\eqno{(A.45)}$$

$$
\tilde{Z}^\alpha=\tilde{\tilde{E}}{}^{\gamma\beta}
\tilde{\Pi}^\alpha_{\beta\gamma}-\tilde{E}^{\gamma\nu\mu}
\tilde{\Pi}^{\beta\rho b}_{\mu\nu\gamma}\tilde{\Pi}^{\alpha a}_{\rho\beta}
\varepsilon_{ab},
\eqno{(A.46)}$$
where

$$\begin{array}{c}
\Pi^{\rho\sigma}_{\mu\nu\alpha}\equiv(\delta^\rho_\mu
\bar{\Pi}^\sigma_{\nu\alpha}(-1)^{\varepsilon_\mu\varepsilon_\nu}-
\delta^\rho_\nu\bar{\Pi}^\sigma_{\mu\alpha}
(-1)^{\tilde{\varepsilon}_\alpha(\varepsilon_\mu+\varepsilon_\nu)})-\\[9pt]
-(\delta^\sigma_\mu\bar{\Pi}^\rho_{\nu\alpha}
(-1)^{\varepsilon_\mu\varepsilon_\nu}
-\delta^\sigma_\nu\bar{\Pi}^\rho_{\mu\alpha}
(-1)^{\tilde{\varepsilon}_\alpha(\varepsilon_\mu+\varepsilon_\nu)})
(-1)^{\varepsilon_\rho\varepsilon_\sigma}+(\imath\hbar)^2
U^{\rho\sigma}_{\mu\nu\alpha},
\end{array}\eqno{(A.47)}$$

$$
\bar{\Pi}^\nu_{\mu\alpha}\equiv-{1\over2}T_\alpha\delta^\nu_\mu
(-1)^{\tilde{\varepsilon}_\alpha\varepsilon_\mu}-\imath\hbar
U^\nu_{\mu\alpha},
\eqno{(A.48)}$$

$$\begin{array}{c}
\tilde{\Pi}^{\beta\rho a}_{\mu\nu\alpha}\equiv(
\bar{\Pi}^{\rho a}_{\mu\nu}\delta^\beta_\alpha+
\bar{\tilde{\Pi}}{}^{\beta a}_{\mu\alpha}\delta^\rho_\nu
(-1)^{\varepsilon_\nu(\tilde{\varepsilon}_\alpha+
\tilde{\varepsilon}_\beta)}-\\[9pt]
-\bar{\tilde{\Pi}}{}^{\beta a}_{\nu\alpha}\delta^\rho_\mu
(-1)^{\varepsilon_\mu(\tilde{\varepsilon}_\alpha+\tilde{\varepsilon}_\beta+
\varepsilon_\nu)})(-1)^{\varepsilon_\mu\tilde{\varepsilon}_\alpha}
+(\imath\hbar)^2\tilde{U}^{a\beta\rho}_{\mu\nu\alpha},
\end{array}\eqno{(A.49)}$$

$$
\bar{\tilde{\Pi}}{}^{\beta a}_{\mu\alpha}\equiv{1\over2}T^a_\mu
\delta^\beta_\alpha-\imath\hbar\tilde{U}^{a\beta}_{\mu\alpha}.
\eqno{(A.50)}$$

\end{document}